# The role of non-scientific factors vis-à-vis the quality of publications in determining their scholarly impact

**Authors:** Giovanni Abramo[1], Ciriaco Andrea D'Angelo[2], Leonardo Grilli[3]

**Affiliations:**

[1] Laboratory for Studies in Research Evaluation, Universitas Mercatorum, Rome, Italy
ORCID: 0000-0003-0731-3635 - giovanni.abramo@unimercatorum.it

[2] University of Rome "Tor Vergata", Dept of Engineering and Management, Rome, Italy
ORCID: 0000-0002-6977-6611 - dangelo@dii.uniroma2.it

[3] University of Florence, Dept of Statistics, Computer Science, Applications "G. Parenti", Florence, Italy
ORCID 0000-0002-3886-7705 - leonardo.grilli@unifi.it

**Abstract**
In the evaluation of scientific publications' impact, the interplay between intrinsic quality and non-scientific factors remains a subject of debate. While peer review traditionally assesses quality, bibliometric techniques gauge scholarly impact. This study investigates the role of non-scientific attributes alongside quality scores from peer review in determining scholarly impact. Leveraging data from the first Italian Research Assessment Exercise (VTR 2001-2003) and Web of Science citations, we analyse the relationship between quality scores, non-scientific factors, and publication short- and long-term impact. Our findings shed light on the significance of non-scientific elements overlooked in peer review, offering policymakers and research management insights in choosing evaluation methodologies. Sections delve into the debate, identify non-scientific influences, detail methodologies, present results, and discuss implications.

## 1. Introduction

In the realm of equivalent cost, does the superior quality product emerge as the top-selling among those fulfilling a specific need? Probably, but not necessarily. Is it solely the product quality that influences consumer choices? Certainly not. The service integrated into or associated with the product, packaging design, distribution channel, promotion, and general marketing are factors that can contribute to enhancing the product's value and/or the consumer's perception of it. The ultimate goal for a company seeking to maximise profits for its shareholders is not only to produce goods or services better than competitors but to sell more (at an equivalent cost). Therefore, the company also invests in activities complementary to those typically aimed at increasing the intrinsic quality of the product.

With the necessary modifications, the taxpayer, i.e., the shareholder of public research institutions (PROs), and consequently, the policymaker and top management overseeing them, should aim at maximising the socio-economic impact of research expenditures rather than solely focusing on the quality of research output. Notably, many of these institutions have established industrial liaison and technology licensing offices (equivalent to the marketing and sales functions of private companies operating in the market) to promote cross-sector knowledge transfer (social impact). Researchers increasingly turn to social media to expedite the speed, reach, and significance of disseminating research results within the scientific community (intra-sector knowledge transfer, i.e., scholarly impact). Similar to consumer goods, empirical evidence for scientific research "products" indicates that, in addition to intrinsic quality, various non-scientific factors play a role in determining their value/impact (Mammola, Piano, Doretto, Caprio, & Chamberlain, 2022; Xie et al., 2019; Tahamtan, Safipour Afshar, & Ahamdzadeh, 2016).

If the impact of research is what Public Research Organizations (PROs) should maximise rather than quality, then why resort to evaluation methods and incentivising systems based on the assessment of quality through peer review of scientific publications? The latest UK Research Evaluation Framework (REF) 2021, the current descendant of the Research Assessment Exercise (RAE) and precursor to other RAEs[1] adopted by an increasing number of countries under various names (e.g., ERA in Australia, PBRF in New Zealand, VQR in Italy, etc.), "is the system for assessing the quality of research in Higher Education Institutions (HEIs) in the UK," where "the primary outcome of the (evaluation) panels' work will be an overall quality profile awarded to each submission." The quality of submitted research outputs in terms of their originality, significance, and rigour constitutes 60 per cent of the overall performance score. In comparison, the social impact "underpinned by excellent research conducted in the submitted unit" accounts for 25 per cent. Additionally, the "vitality and sustainability" of the research environment contribute 15 per cent.[2] The introduction of social impact evaluation in RAEs is relatively recent and conducted through the analysis of a very limited number of case studies. In any case, the assessment of the quality of research output continues to play a primary role both in the final evaluation and in the performance-based allocation of resources to institutions. These exercises often use a combination of methods, but peer review plays a central role in determining the quality of research outputs and, consequently, the overall research performance of institutions and individuals.

---

[1] Henceforth, we will use the acronym 'RAEs' to generically refer to national research assessment exercises.
[2] The above phrases in quotes are extracted from REF 2021 – Panel criteria and working methods, https://www.ref.ac.uk/publications-and-reports/panel-criteria-and-working-methods-201902/

The debate on which of the two approaches is preferable for research evaluation has recently been reignited by the Coalition for Advancing Research Assessment (CoARA) initiative. CoARA advocates that research assessment should be primarily based on qualitative judgment, with peer review playing a central role[3] (Rushworth, 2023; Torres-Salinas, Arroyo-Machado, & Robinson-Garcia, 2023). It is unequivocal that both approaches have pros and cons. Still, it is important to acknowledge that the two methods measure different attributes of research one being the quality of scientific output and the other its scholarly impact. The potential difference between the two should be determined by non-scientific factors associated with the publication, which will be the subject of the present study.

This study aims to ascertain to what extent non-scientific factors contribute to determining, in addition to intrinsic quality, the scholarly impact of a research product. To achieve this, we assume that peer review competently measures quality, and citation-based metrics measure the scholarly impact of research products despite the respective limitations of the two methodologies extensively dissected in the literature (Gingras, 2016; Lee, Sugimoto, Zhang, & Cronin, 2013).

We leverage the knowledge of the quality scores attributed by reviewers to the publications submitted for evaluation in the first Italian RAE, named VTR 2001-2003. For each such publication indexed in the Web of Science (WoS), we measure its citation-based impact. Subsequently, we identify the non-scientific attributes of each publication. Finally, we fit a statistical model to analyse the relationship between the impact of the publication and the quality score assigned by reviewers, controlling for the non-scientific factors. This allows us to gain insight into the weight of non-scientific factors that reviewers may not capture in determining the scholarly impact of a publication.

The results should interest policymakers and research institution management, who must decide whether to measure the quality of research output through peer review or academic impact through bibliometric techniques.

The following section presents the main insights from the debate on peer-review vs bibliometric approaches to assess research. In Section 3, we examine the non-scientific factors that could influence the scholarly impact of publications. In Section 4, we provide details on the data and methods used to measure their relative contribution to impact. In Section 5, we present the results, while in Section 6, we discuss them and draw conclusions.

## 2. Quality or impact, peer review or bibliometrics?

For a research product to impact the scientific community, it must be utilised (OECD/Eurostat, 2018). To ensure this, it cannot remain tacit but must be encoded in written form beforehand to facilitate its dissemination across all potential users. The most commonly adopted written form by scholars is the article published in scientific journals. The utilisation of the knowledge embedded in the publication is primarily determined by its quality, encompassing originality, significance, and rigour (Jabbour, Jabbour, & de Oliveira, 2013; Patterson & Harris 2009), as well as other non-scientific attributes (Mammola et al., 2022; Xie et al., 2019; Tahamtan, Safipour Afshar, & Ahamdzadeh, 2016). The contribution of these non-scientific factors may allow a publication of lower quality to have a greater impact than a qualitatively superior one. Therefore, if the intention is to measure impact, evaluating quality alone could lead to distorted results. While bibliometric techniques directly measure a publication's scholarly impact, determined by its quality and non-scientific factors to some extent, the peer-review process is intended to judge its quality.

---

[3] https://coara.eu/

The pros and cons of each approach have been extensively debated in the literature. The major limitation of peer evaluation is the subjectivity in assessments, as highlighted by the not infrequent disagreements among peers (Schroter, Weber, Loder, Wilkinson, & Kirkham, 2022; Kirman, Simon, & Hays, 2019; Bertocchi, Gambardella, Jappelli, Nappi, & Peracchi, 2015), especially in the case of interdisciplinary works (Thelwall et al., 2023a). Subjectivity occurs not only in the evaluation of the research product but also in the upstream phase of selecting peers (Moxham & Anderson, 1992; Horrobin, 1990) and in the selection of products to be subjected to evaluation (Abramo, D'Angelo, & Di Costa, 2014). Other limitations include potential conflicts of interest, the natural tendency to more favourably evaluate authors with a higher reputation or affiliated with more prestigious institutions, the difficulty of contextualising the judgment on the work to be evaluated in the state of the art at the time of execution, the challenge of identifying quality reviewers as the number of works to be evaluated increases (Abramo, D'Angelo, & Viel, 2013), and last but not least, the costs and time involved in large-scale evaluations, RAEs. Notably, RAEs that adopt peer evaluation base the comparative assessment of institutions on a limited number of total research products to reduce costs and time, albeit at the expense of inevitable distortions in rankings (Abramo, D'Angelo, & Viel, 2010).

Citation-based metrics of the scholarly impact of a publication presuppose that a citation represents a recognition of the influence of the cited publication on the citing one (normative theory of citing) (Mulkay, 1976; Bloor, 1976; Merton, 1973). This assumption is strongly opposed by social constructivists who, on the contrary, believe that persuasion is the primary reason for citing, and therefore, citation-based metrics are not suitable for measuring the impact of scientific work (Brooks, 1985; MacRoberts & MacRoberts, 1984; Knorr-Cetina, 1981; Latour & Woolgar, 1979). The literature has extensively discussed this opposition (Tahamtan & Bornmann, 2018). However, the fact that there is a significant correlation between quality as judged by peers and scholarly impact as measured by citation metrics, both perceived and empirically observed (Jabbour, Jabbour, & de Oliveira, 2013; Patterson & Harris 2009), supports the normative theory of citing.

This does not mean that citations as an impact indicator are free from limitations. In fact, citations do not always reflect quality, as a work can be cited to demonstrate its faults rather than its merits (negative citations). However, this rare event generally occurs soon after publication and should not disrupt the analyses (Pendlebury, 2009). Citations can also be manipulated by authors for their own benefit (excessive recourse to self-references and cross-citations) or by editors who, in some cases, exert pressure on authors to cite works already published in their journals to increase the journals' impact indicators (Pichappan & Sarasvady, 2002). Another limitation is the "delayed recognition" phenomenon, which sometimes affects more mature works (Garfield, 1980; van Raan, 2004; Ke, Ferrara, Radicchi, & Flammini, 2015). Although it has been demonstrated that cases of delayed recognition are quite rare, their effects can still be mitigated by introducing other variables beyond early citations that improve the predictive power of the latter (Xia, Li, & Li, 2023; Abramo, D'Angelo, & Felici, 2019). Finally, the use of citations requires reliance on bibliographic repertories, which do not index all publications and, in disciplines such as arts and humanities, their coverage is insufficient to provide a robust representation of research output (Aksnes & Sivertsen, 2019; Archambault et al., 2006; Moed, 2005).

The debate on which of the two approaches is more appropriate is still open and will likely remain so for a long time. There is no definitive answer: the choice between the two, either individually or in combination, will depend on the measurement goal, context, measurement scale, data availability, and resources and time constraints.

In the next section, we will analyse the non-scientific attributes of a scientific publication that, together with its intrinsic quality, determine its scholarly impact.

## 3. Non-scientific factors affecting publications impact

A comprehensive body of literature has emerged on the non-scientific factors influencing publication impact, as evidenced in the review by Tahamtan, Safipour Afshar, and Ahamdzadeh (2016). Xie et al. (2019) identified 66 factors possibly associated with impact.

These factors can be classified into two main categories: those external to the manuscript and those intrinsic to the manuscript. Among the former, notable elements include i) knowledge distribution channels such as the prestige level of the publishing journal (Mammola, Fontaneto, Martínez, & Chichorro, 2021; Bornmann & Leydesdorf, 2015; Stegehuis, Litvak, & Waltman, 2015) and the type of access (open access, or OA, vs non-open access, or non-OA) to the publication by a potential reader (Yu, Meng, Qin, Shen, & Hua, 2022; Langham-Putrow, Bakker, & Riegelman, 2021; Piwowar, et al., 2018; Wang, Liu, Mao, & Fang, 2015; Gargouri, et al., 2010; Antelman, 2004); and ii) communication initiatives in social media (i.e., blogs, Twitter, Facebook, pre-prints) undertaken by the authors to increase the visibility of the manuscript (Özkent, 2022).

Intrinsic factors within the manuscript can be further categorised into three groups based on the part of the manuscript they pertain to i) features of the byline, ii) features of the body of the manuscript, and iii) features of the reference list.

Regarding the features of the byline, factors associated with impact include i) the length of the author list (Thelwall et al., 2023b; Talaat & Gamel, 2022; Fox, Paine, & Sauterey, 2016; Abramo & D'Angelo, 2015; Didegah & Thelwall, 2013; Wuchty, Jones, & Uzzi, 2007); ii) the authors' academic influence and collaboration network (Hurley, Ogier, & Torvik, 2013; Mammola, Piano, Doretto, Caprio, & Chamberlain, 2022); iii) authors' gender or other personal features (Abramo, Aksnes, & D'Angelo, 2021; Andersen, Schneider, Jagsi, & Nielsen, 2019; Duch et al., 2012; Aksnes, Rorstad, Piro, & Sivertsen, 2011; Larivière, Vignola-Gagné, Villeneuve, Gelinas, & Gingras, 2011; Symonds, Gemmell, Braisher, Gorringe, & Elgar, 2006); iv) the number of institutions collaborating (Sanflippo, Hewitt, & Mackey, 2018; Narin & Whitlow, 1990); v) the number of countries involved (Glänzel & De Lange, 2002).

Moving to the features of the body of the text, factors linked to impact include i) document types (e.g., articles, reviews, proceedings papers, books, etc.), which are differently associated with impact and speed of impact (Wang, Song, & Barabási, 2013); ii) linguistic attributes of the manuscript (including title and abstract) such as readability (Ante, 2022; Heßler, Ziegler, 2022; Rossi & Brand, 2020; Stremersch, Camacho,

Vanneste, & Verniers, 2015; Didegah & Thelwall, 2013; Walters, 2006); iii) the manuscript's length (Xie, Gong, Cheng, & Ke, 2019; Elgendi, 2019; Fox, Paine, & Sauterey, 2016; Ball, 2008); iv) the degree of interdisciplinarity (Chen, Arsenault, & Larivière, 2015; Yegros-Yegros et al., 2015); v) popularity and interest of the subject (Peng & Zhu, 2012); vi) the discipline the manuscript falls under (Larivière & Gingras, 2010; Levitt & Thelwall, 2008); and vii) research fundings (Rigby, 2013).

Finally, concerning the reference list, factors associated with impact include i) the length of the reference list (Mammola, Fontaneto, Martínez, & Chichorro, 2021; Fox, Paine, & Sauterey, 2016); ii) the impact of the cited works (Sivadas & Johnson, 2015; Jiang, He, & Ni, 2013); iii) the incidence of more recent cited publications (Liu, Chen, Liu, Bu, & Gu, 2022; Mammola, Fontaneto, Martínez, & Chichorro, 2021); iv) the number of cited fields and their cognitive distance (Wang, Thijs, & Glänzel, 2015).

The non-scientific traits of a manuscript associated with its future impact are numerous, but not all of them are easily measurable in large-scale analyses.

## 4. Data and methods

### 4.1 Data

To better understand the methodology employed to address our research query, it is essential to delve into the VTR 2001-2003, which adopted a peer-review assessment. The primary objective of the VTR was to evaluate the research conducted by Italian universities and public research institutions during the specified timeframe. Each of the 102 institutions under scrutiny, encompassing 64,000 researchers, was tasked with submitting their research products during this timeframe. A restriction was imposed to ensure that the number of products did not exceed 50%[4] of the full-time-equivalent (FTE) research staff of each evaluated university. Reviewers were entrusted with evaluating a comprehensive pool of 17,329 research products. Following the conclusion of the peer review process, each research product received a final judgment, expressed on a four-point rating scale: Excellent (E) = 1, denoting products that met the top 20% of international standards; Good (G) = 0.8 for those falling between 80% and 60%; Acceptable (A) = 0.6 for products scoring between 60% and 40%; and Limited (L) = 0.2 for products falling below the 40% threshold. Each product selected by universities and sent to the agency in charge of the evaluation was classified into a particular subject area (DAs, 14 in all). For the purposes of this paper, we have limited the analysis only to products (9,225 in all) classified in 7 of such areas and, more precisely: 1 - Mathematics and computer science; 2 - Physics; 3 - Chemistry; 4 - Earth sciences; 5 - Biology; 7 - Agricultural and veterinary sciences; 8 - Civil engineering; 9 - Industrial and information engineering. Notably, these cover all Sciences except Medicine.

These products are mainly journal articles (72%), but there are also books and book chapters (23%), patents (2%), and the remaining 3% of miscellaneous items.

In certain instances, co-authors from distinct institutions submitted identical products (in a very limited number of cases, in distinct DAs). Among the submitted products, 8,086 were scientific publications indexed in the Web of Science (WoS). Some of them were excluded since i) exhibited publication dates outside the 2001-2003 period; ii) were hosted in a source lacking *impact factor* (IF); or were assigned to different DA panels (as anticipated above), and received different evaluation scores. After eliminating all the aforementioned cases, the finalised dataset comprised 7,305 publications, whose breakdown by DAs is shown in Table 1. This corresponds to 11.8% of the WoS-indexed scientific production of all Italian academics within those DAs, ranging from a minimum of 7.2% in Industrial and information engineering to a maximum of 27.3% in Earth sciences.

---

[4] To account for time devoted to teaching activities, 1 professor equals 0.5 FTE.

**Table 1. Number of Italian publications in the dataset by disciplinary area**

| Disciplinary area | Publications in the dataset (a) | Total Italian publications** (b) | a/b |
|---|---|---|---|
| 1 - Mathematics and computer science | 727 | 6,258 | 11.5% |
| 2 - Physics | 1,543 | 12,414 | 12.4% |
| 3 - Chemistry | 1,001 | 12,958 | 7.7% |
| 4 - Earth sciences | 581 | 2,126 | 27.3% |
| 5 - Biology | 1,504 | 14,545 | 10.3% |
| 7 - Agricultural and veterinary sciences | 667 | 4,053 | 16.3% |
| 8 - Civil engineering | 350 | 2,323 | 14.7% |
| 9 - Industrial and information engineering | 993 | 13,706 | 7.2% |
| Total* | 7,305 | 61,592 | 11.8% |

*\* The figures on the last line do not match the column total due to products falling into multiple DAs being counted more than once.*
*\*\* Total 2001-2003 WoS publications authored by professors in each DA.*

## 4.2 Methods

### 4.2.1 Variables

For the analysis, we focus on "journal articles" only, which are the vast majority of the research products in the dataset (6,889 articles out of 7,305). To assess the impact of the articles, we use WoS bibliometric data. Specifically, we calculate the normalised scholarly impact of publication $i$ as its citations accrued up to 31/12/2022, normalised to the reference distribution, i.e. divided by the average number of citations (counted at the same date) received by all WoS publications classified in the same subject category (SC)[5,6] and indexed in the same year of publication $i$. Having considered a fixed citation count date, the citation window used to assess the impact of publications varies from a minimum of 19 years to a maximum of 21. In all cases, these are extremely long, which ensures a reliable measurement of long-term scholarly impact. For ease of reading, in the following, we call the scholarly impact simply "impact", and, being our target variable, we will denote it by Y. In fact, as will be clearer below, we will also consider the "short-term impact" of each publication, measured exactly as indicated above, but considering citations received on 31/12/2005, i.e., with the same time citation window available to the evaluators in the VTR.

We assume that the impact depends mainly on "quality", denoted with Z, for the measurement of which we rely on the final judgment expressed by reviewers; hence, quality is expressed through the four-point rating scale already described above (1 for "excellent products"; 0.8 for "good" ones; 0.6 for "acceptable" and 0.2 for "limited").

---

[5] The WoS classification scheme involves 255 SCs in all.

[6] The SC of a publication corresponds to that of the journal where it is published. For publications in multidisciplinary journals (multiple SCs) the scaling factor is calculated as the average of the standardized values for each subject category.

On the other hand, as the literature suggests, we also assume that impact depends on several non-scientific factors ($X_1, \ldots, X_p$). We group them into three sets of features.

Features of the byline:

- The number of authors of the publication.
- The average impact of their 2001-2003 publications (measured as the Y).
- A dummy variable for the presence of an English mother tongue author for modelling the possible "linguistic advantage".
- The share of female co-authors.
- The number of institutions and countries in the affiliations list. The strong correlation between these two variables leads us to exclude both in favour of a single dummy ("foreign") equal to 1 when the address list contains more than one country (0, otherwise).

Features related to the publication's content and venue:

- The open-access character, expressed by a single dummy equal to 1 for Green, Hybrid, and Gold OA-tagged publications.
- The length of the publication expressed by the number of pages.
- Impact_factor of the hosting source, extracted from the Journal Citation Report 2004 edition and normalised to the reference distribution, i.e., divided by the average impact factor of all sources in the same subject category.
- The degree of interdisciplinarity of the publication measured by the share of cited papers in the bibliography falling in SCs other than the dominant one in the article reference list (Abramo, D'Angelo, & Zhang, 2018).

Features related to the publication's bibliography:

- The length of the reference list (number of cited references).
- The share of references indexed in WoS signalling the extent of recourse to "qualified" literature.
- The share of self-citations in the reference list.
- The average age of cited publications.
- Their average normalised impact (measured as the outcome Y).

Some additional explanations deserve to be given on how the features related to the byline were measured. Specifically, each author of the publications in the dataset has been "disambiguated" using the algorithm proposed by Caron and van Eck (2014). This enables us to attribute their gender and nationality, survey their past scientific production, and measure the average impact of their 2001-2003 publications.

We dropped 220 articles with missing values on the authors' average impact and 223 with outlying values in the number of authors, pages, or references. Hence, the final dataset for the analysis includes 6,446 articles.

Table 2 shows summary statistics of the variables of the dataset. For better readability, we omit all variables found to be not statistically significant at $p$=0.05 in terms of their effect on the response (Y), both here and in the following.

*Table 2. Summary statistics of the variables of the dataset (6,446 articles)*

| Variable | Mean | Std dev. | Min | Max |
|---|---|---|---|---|
| *Outcome* | | | | |
|   Impact | 1.545 | 2.861 | 0 | 91.917 |
| *Quality* | | | | |
|   Peers' score | | | | |
|     Limited (0.2) | 2.2% | | | |
|     Acceptable (0.6) | 16.6% | | | |
|     Good (0.8) | 49.0% | | | |
|     Excellent (1) | 32.1% | | | |
|   Short-term impact | 1.280 | 1.739 | 0 | 29.564 |
| *Byline* | | | | |
|   N. authors | 5.508 | 6.718 | 1 | 99 |
|   Avg authors impact | 0.967 | 0.829 | 0 | 25.845 |
|   Foreign | 0.369 | 0.483 | 0 | 1 |
| *Content and venue* | | | | |
|   Open Access | 0.318 | 0.466 | 0 | 1 |
|   N. pages | 10.955 | 8.355 | 1 | 95 |
|   Journal IF | 33.119 | 18.25 | 2 | 196 |
| *Bibliography* | | | | |
|   N. references | 2.215 | 2.02 | 0 | 15.792 |
|   % references in WoS | 68.711 | 22.328 | 2.941 | 100 |
|   % self-cites | 22.188 | 19.118 | 0 | 100 |
|   Cited articles age | 7.021 | 2.644 | 0 | 21 |
|   Cited articles avg impact | 10.285 | 22.26 | 0.035 | 520.765 |

### 4.2.2 The statistical model

We exploit a linear random effects model to evaluate the extent of the association of the above-mentioned non-scientific features via-à-vis the quality of a given publication with its scholarly impact (e.g., Rabe-Hesketh & Skrondal, 2022). Specifically, for an article $i$ published in journal $j$, the model is:

$$Y_{ij} = \alpha + \delta Z_{ij} + \beta_1 X_{1ij} + \cdots + \beta_p X_{pij} + u_j + e_{ij} \qquad [1]$$

where $Y_{ij}$ is the outcome variable, i.e., the article's (normalised) impact, $Z_{ij}$ is a measure of the article's quality and $X_{1ij}, \dots, X_{pij}$ are the article's non-scientific factors.

Model [1] includes a random effect $u_j$ representing the unobserved factors for journal $j$ and a residual error $e_{ij}$. Both $u_j$ and $e_{ij}$ are assumed to be normal random variables with zero mean and unknown standard deviations denoted with $\sigma_u$ and $\sigma_e$, respectively. The model is fitted using the *xtreg* command of Stata 18 with the *re* option (StataCorp., 2023).

A preliminary analysis showed that the relationships are linear on the logarithmic scale for all "impact" variables with right-skewed distributions. For this reason, those variables are log-transformed. Values equal to zero are replaced with the minimum positive value before computing the logarithm.

### 5. Results

We fitted three models, each including all the relevant non-scientific factors in Table 2. The response variable is the logarithm of the article's impact. The explanatory variables based on a measure of impact are also logarithmic. The first model includes only non-scientific factors as predictors, the second model adds the "quality" of the article in terms of the score assigned by the peers, and the third model substitutes the quality scores with the short-term impact as measured by the early citations. The results are reported in Table 3. The R-squared, a measure of predictive ability, is 0.248 when considering only non-scientific factors. It increases slightly to 0.260 when including the score assigned by peers and substantially to 0.485 when adding early citations. Therefore, the value added by the peers is marginal and much smaller than the contribution given by an easily and cheaply calculated bibliometric indicator of short-term impact.

All explanatory variables are statistically significant at the 0.01 level, except for open access in all models and the number of authors in the third one. Moreover, in the second model, the categories 0.2 and 0.6 of the score assigned by peers do not achieve statistical significance ($p$-values equal to 0.8909 and 0.0650, respectively): This means that, when controlling for non-scientific factors, articles with scores of 0.2 and 0.6 do not exhibit statistically different impacts compared to those with a score of 0.8 (baseline). However, the top score, namely 1, has a highly significant effect ($p$-value < 0.0001), meaning that reviewers better identify excellent rather than poor papers. Specifically, an article scoring 1 by peers has, other things being equal, a long-term scholarly impact about 28% greater than an article scoring 0.8. On the other hand, the coefficient of the short-term impact is an elasticity measure; thus, a 1% increase in the early citations is associated with a 0.446% increase in the long-term impact.

Non-scientific factors consistently influence the outcome of all models. Their effects tend to be smaller in magnitude in the model with the early citations. The reason is that the impact measured by early citations is a mediator for the long-term impact, absorbing part of the effects; nonetheless, it is remarkable that most non-scientific factors are relevant, even controlling for the early citations. Specifically, as for the byline, the number of authors is not statistically significant, whereas an increase of 1% in the average author's impact is associated with a 0.246% increase in the long-term impact, and the presence of a foreign author is associated with a 7.2% increase. As for content and venue, open access is not statistically significant, whereas one more page length is associated with 1.1% more impact, and an increase of 1% in the journal IF corresponds to 0.267% more impact. As for features related to the publication's bibliography, we found different patterns: the effect on the long-term impact is positive for the number of references (+0.4% for one additional reference) and the average impact of cited articles (+0.101% for 1% more average impact). On the other hand, the relationship is negative for the cited articles' age (-2.3% for one additional year), the percentage of self-cites (-0.4% for one additional percentage point), and the percentage of cited articles in WoS (-0.4% for one additional point). The estimated coefficients generally align with expectations, except for the negative coefficient associated with the percentage of cited articles in WoS. This unexpected result may warrant further investigation. However, it should be noted that this effect is adjusted for the other explanatory variables, particularly the bibliometric ones.

Considering the residual variances, it is worth noting that inserting the peers' score in the second model has negligible consequences. On the other hand, inserting the early impact in the third model leaves the journal-level variance unchanged while substantially reducing the article-level variance; thus, the fraction of residual variance due to journals rises from 19.6% to 27.3%. It is remarkable that in the third model, 27.3% of the residual variance is at the journal level, i.e., more than one-fourth of the variance due to unmeasured factors is attributable to the journal. This implies a residual correlation of 0.273 among articles published in the same journal, even controlling for the journal IF. Therefore, the journal plays an important role in determining the long-term impact beyond the value of its IF.

The analysis has been replicated at the disciplinary area level of the articles. Table 4 shows the R-squared for the three fitted models. The overall pattern is confirmed in all areas: once the non-scientific factors are considered, the score assigned by peers barely improves the prediction of the long-term impact, whereas the short-term impact provides a substantial improvement, especially in Physics and Biology.

*Table 3. Point estimates (standard errors in parenthesis) for three linear random effects models for the logarithm of the article's impact controlling for non-scientific factors (6,446 articles in 1,211 journals)*

|  | Quality/Impact | | |
|---|---|---|---|
|  | None | Peer review | Short-term impact |
| R-squared | 0.248 | 0.260 | 0.485 |
| Intercept | -0.848 (0.095) | 0.095 (-0.792) | -0.792 (0.096) |
| *Byline* | | | |
| N. authors | 0.007 (0.002) | 0.007 (0.002) | 0.003 (0.002) |
| Avg authors' impact (log) | 0.381 (0.018) | 0.364 (0.018) | 0.246 (0.015) |
| Foreign | 0.127 (0.027) | 0.112 (0.027) | 0.072 (0.023) |
| *Content and venue* | | | |
| Open Access | 0.076 (0.035) | 0.058 (0.035) | 0.019 (0.030) |
| N. pages | 0.014 (0.002) | 0.011 (0.002) | 0.011 (0.002) |
| Journal IF (log) | 0.668 (0.060) | 0.593 (0.061) | 0.267 (0.056) |
| *Bibliography* | | | |
| N. references | 0.008 (0.001) | 0.008 (0.001) | 0.004 (0.001) |
| % references in WoS | -0.004 (0.001) | -0.004 (0.001) | -0.004 (0.001) |
| % self-cites | -0.003 (0.001) | -0.004 (0.001) | -0.004 (0.001) |
| Cited articles age | -0.062 (0.005) | -0.061 (0.005) | -0.023 (0.004) |
| Cited articles avg impact (log) | 0.165 (0.013) | 0.163 (0.013) | 0.101 (0.011) |
| *Quality* | | | |
| Peers' score | | | |
| Limited (0.2) | | -0.012 (0.085) | |
| Acceptable (0.6) | | -0.066 (0.036) | |
| Good (0.8) – baseline | | --- | |
| Excellent (1) | | 0.280 (0.030) | |
| Early citations (log) | | | 0.446 (0.008) |
| *Residual variances* | | | |
| Level 2 (journal) | 0.212 | 0.211 | 0.217 |
| Level 1 (article) | 0.869 | 0.854 | 0.579 |
| Fraction of level 2 variance | 0.196 | 0.198 | 0.273 |

*Table 4. Sample size and values of the R-squared for three linear random effects models for the logarithm of the article's impact controlling for non-scientific factors, by disciplinary area*

| | | Quality/Impact | | |
|---|---|---|---|---|
| Disciplinary area | Obs | None | Peer review | Short-term impact |
| 1 - Mathematics and computer science | 683 | 0.29 | 0.30 | 0.41 |
| 2 - Physics | 1184 | 0.33 | 0.37 | 0.66 |
| 3 - Chemistry | 912 | 0.24 | 0.25 | 0.52 |
| 4 - Earth sciences | 536 | 0.23 | 0.24 | 0.54 |
| 5 - Biology | 1331 | 0.27 | 0.30 | 0.62 |
| 7 - Agricultural and veterinary sciences | 627 | 0.25 | 0.25 | 0.42 |
| 8 - Civil engineering | 307 | 0.21 | 0.23 | 0.42 |
| 9 - Industrial and information engineering | 921 | 0.21 | 0.22 | 0.38 |
| Overall | 6446 | 0.25 | 0.26 | 0.49 |

## 6. Discussion and conclusions

The prevailing view among the majority of research assessment scholars and practitioners is that peer-review evaluation stands as the gold standard, while citation-based evaluation serves as a quicker and more economical surrogate. Consequently, considerable scholarly attention has been directed towards assessing how effectively bibliometric evaluation approximates peer review, with potential for substitution or complementary use across various domains: i) individual scientific publications (Bertocchi, Gambardella, Jappelli, Nappi, & Peracchi, 2015; Bornmann & Leydesdorff, 2013); ii) individual researchers (Vieira & Gomes, 2018; Cabezas-Clavijo, Robinson-García, Escabias, & Jiménez-Contreras, 2013), and iii) research institutions (Pride & Knoth, 2018; Franceschet & Costantini, 2011).

However, the underlying rationale for the presumption of peer review's superiority over evaluative bibliometrics remains elusive.

On the empirical front, Abramo, D'Angelo, and Reale (2019) presented evidence that early citations offer better predictions of the long-term scholarly impact of scientific publications compared to peer-review quality scores. The present study reinforces these findings, even after factoring in the influence of non-scientific factors on impact.

Similarly, theoretical foundations are also wanting. In recent years, policymakers have rightly shifted their focus towards assessing the social impact of research rather than merely its quality. While research output quality serves as a means to an end, the ultimate goal of research activity invariably remains its societal impact. This impact is influenced partly by research institutions and partly by the capacity of both the industrial and public sectors to incorporate research findings into enhanced products and processes. Concerning institutional responsibilities, a clear division between production and dissemination - between researchers and technology licensing offices - is essential. Researchers must not only generate quality results but also ensure their swift dissemination within the scholarly community, utilising non-scientific factors responsibly to enhance scholarly impact.

Recent national evaluation initiatives have partly embraced these new priorities, mandating institutions to demonstrate the societal impact of their research by submitting relevant case studies. However, by retaining peer-review evaluation of publications as the primary metric, these assessments continue to prioritise research quality over scholarly impact. Peers, often experts in specific fields, face challenges in assessing the broader scientific implications of research beyond their expertise. Recent studies reveal that citations to publications predominantly originate from diverse domains (Abramo & D'Angelo, 2024).

While the quality of scientific products typically plays a significant role in scholars' choices, other factors also exert influence. As demonstrated in this study concerning scientific articles, factors such as the reputation of authors and the publication venue, alongside quality, shape their utilisation by the scientific community for advancing knowledge. Additionally, factors like the average impact of cited works and multinational authorships contribute to a lesser extent. Embracing bibliometric methods, which effectively measure scholars' citations as "consumer purchases," could advance policymakers' shared objectives. Replacing "evaluation of research quality" with "evaluation of research impact" in technical parlance would foster greater clarity among professionals and stakeholders regarding evaluation objectives.

Given these research findings and considerations, the recent initiative by CoARA towards predominantly peer-review evaluation systems raises several perplexities and concerns that policymakers should carefully address.